\documentclass[final,3p,times,twocolumn]{elsarticle}

\usepackage{lineno,hyperref}
\usepackage{amssymb}
\usepackage{graphicx}
\usepackage{floatrow}
\usepackage{upgreek}
\usepackage{lineno}

\modulolinenumbers[1]

\journal{Nuclear Instruments and Methods in Physics Research Section A}

\begin{document}
\begin{frontmatter}

\title{A new technique for elucidating $\beta$-decay schemes which involve daughter nuclei with very low energy excited states}

\author[IP]{M.~Venhart\corref{cor1}}
\ead{martin.venhart@savba.sk}
\cortext[cor1]{Corresponding author}
\author[GA]{J.~L.~Wood}
\author[IP]{A.~J.~Boston}
\author[MAN,IKS]{T.~E.~Cocolios}
\author[OLL]{L.~J.~Harkness-Brennan}
\author[OLL]{R.-D.~Herzberg}
\author[OLL]{D.~T.~Joss}
\author[OLL]{D.~S.~Judson}
\author[IP]{J.~Kliman}
\author[IP]{V.~Matou\v{s}ek}
\author[FEI]{\v{S}.~Moty\v{c}\'ak}
\author[OLL]{R.~D.~Page}
\author[OLL]{A.~Patel}
\author[IP]{K.~Petr\'{i}k}
\author[IP]{M.~Sedl\'{a}k}
\author[IP]{M.~Veselsk\'{y}}

\address[IP]{Institute of Physics, Slovak Academy of Sciences, SK-84511 Bratislava, Slovakia}
\address[GA]{Department of Physics, Georgia Institute of Technology, Atlanta, GA 30332, USA}
\address[MAN]{School of Physics and Astronomy, The University of Manchester, Manchester M13 9PL, United Kingdom}
\address[IKS]{KU Leuven, Instituut voor Kern- en Stralingsfysica, B-3001 Leuven, Belgium}
\address[OLL]{Oliver Lodge Laboratory, University of Liverpool, Liverpool, L69 7ZE, UK}
\address[FEI]{Faculty of Electrical Engineering and Information Technology, Slovak University of Technology, SK-812 19 Bratislava, Slovakia}

\begin{abstract}
A new technique of elucidating $\beta$-decay schemes of isotopes with large density of states at low excitation energies has been developed, in which a Broad Energy Germanium (BEGe) detector is used in conjunction with coaxial hyper-pure germanium detectors.  The power of this technique has been demonstrated on the example of $^{183}$Hg decay. Mass-separated samples of $^{183}$Hg were produced by a deposition of the low-energy radioactive-ion beam delivered by the ISOLDE facility at CERN.  The excellent energy resolution of the BEGe detector allowed $\gamma$ rays energies to be determined with a precision of a few tens of electronvolts, which was sufficient for the analysis of the Rydberg-Ritz combinations in the level scheme. The timestamped structure of the data was used for unambiguous separation of $\gamma$ rays arising from the decay of $^{183}$Hg from those due to the daughter decays.
\end{abstract}

\date{\today}


\end{frontmatter}

\section{Introduction}


The $\beta$~decay process plays a fundamental role in studies of nuclear structure.  The process is defined by an initial state, a ground state or an isomer of the parent isotope, which ideally has an accurately known energy (mass), spin and parity.  To study this process, modern experimental methods are used to produce very pure sources using isotope separation techniques.  Atoms (ions) can be separated by mass either electromagnetically or by using a laser induced ionisation step~\cite{Blu13}, with the latter also able to separate isomers from ground states. The decay process results in the population of states in the daughter nucleus with spins close to that of the parent nucleus. Elucidation of the excited states of the daughter nucleus proceeds via $\gamma$-ray and conversion-electron spectroscopy~\cite{Rup98,Pap88}. Energies of excited states are deduced via energy sums and differences of the observed radiations.  Spins and parities of excited states are deduced via multipolarities of the observed radiations, in reference to the (known) ground-state spin and parity of the daughter nucleus.  Angular correlation or angular distribution data for $\gamma$ rays can also be used to determine spins and multipolarities.

Radioactive decay plays a leading role in the modern study of nuclear structure, through the experimental investigation of the thousands of nuclei that lie far from stability.  The challenge is in the complexity of these studies, especially when the nuclei involved possess odd mass or, even more challenging, when the daughter is an odd-odd nucleus, due to the high level density.  Herein we address the further challenge where a high level density occurs at very low excitation energy.

The occurrence of a high level density at low excitation energy results in major challenges for observing the de-exciting transitions~\cite{Rup98}. Generally, low-energy transitions proceed preferentially by internal conversion. High-resolution conversion electron spectroscopy~\cite{Zga16} is technically very demanding and only a few research groups have pursued this technique. Further, the use of coincidence spectroscopy to reliably sequence decay paths can be difficult or impossible because of isomerism (coincidence delay) occurring for some of the low-lying excited states. In the face of such challenges, only the Rydberg-Ritz combination principle~\cite{Rit08} provides a means for elucidating a decay scheme, thus providing reliable excited state information for the daughter nucleus.  Generally, the complexity of odd-mass decay schemes (many hundreds of $\gamma$ rays) renders the Rydberg-Ritz technique too ambiguous to be useful, unless $\gamma$-ray energies are measured to a precision better than 50\,eV.  We have therefore developed and applied a new experimental technique for these studies.

In this work, a Broad Energy Germanium detector (BEGe)~\cite{Har14} has been used in combination with standard coaxial germanium detectors in the study of the $^{183}$Hg~$\rightarrow$~$^{183}$Au decay scheme using mass separated sources from the ISOLDE facility at CERN. This decay possesses a feature that occurs commonly when the $Q$ value of the $\beta$ decay is large, namely the concentration of the $\beta$ decay process to a few highly excited states in the daughter nucleus. These states can decay to many low-lying states, primarily by $\gamma$-ray emission. Using a combination of the high resolution of the BEGe detector and $\gamma$-$\gamma$ coincidence spectroscopy, energy differences of decay sequences reveal low-lying excited states even where the transitions between the low-lying excited states are not observed. Further, we operated the BEGe detector at a gain of 27\,eV/ch so that energies accurate to $\pm$\,10\,eV could be obtained for most of the observed $\gamma$ rays.  

\section{Experimental details}

The experiment was performed at the ISOLDE facility located at CERN. It uses a pulsed beam of protons with an energy of 1.4\,GeV and an average intensity of 1.5\,$\upmu$A for production of radioactive ion beams. The $^{183}$Hg species were produced by spallation reactions in a molten lead target. The reaction products were diffused out of the target using high temperature, ionised with a plasma ion source and extracted with a 30\,kV electric potential. After extraction from the ion source, the beam was separated employing the General Purpose Separator of the ISOLDE facility, which has one analysing magnet~\cite{Kug00}. 

The mass-separated radioactive-ion beam of $^{183}$Hg with an energy of 30\,keV was then delivered to the TATRA tape transportation system~\cite{Mat15}. The samples were created by a deposition of the beam on a tape made of amorphous metal. The activity was collected for a period of 1\,s, after which the sample was transported into the measurement position where $\gamma$ rays following the radioactive decay were detected by an array of three different High Purity Germanium (HPGe) detectors. A BE2020 type BEGe \cite{Har14} detector, which has a non-bulletized disc shape with an active diameter of 51\,mm and a thickness of 20\,mm was used to detect $\gamma$ rays within the 40\,-\,980\,keV range. This detector has excellent energy resolution and a low-energy detection limit, relative to coaxial detectors, due to the electrode configuration employed in BEGe detectors and a fabrication process that optimises the charge collection times across a wide energy range. Two additional p-type coaxial detectors with relative efficiency of 70\,\% were used to detect $\gamma$ rays up to 2.5\,MeV. Both coaxial detectors were mounted perpendicularly to the BEGe detector. A source-to-detector distance of 5\,cm was used for all three detectors. After 30\,s of data collection, a new sample was made and the process was repeated.

\begin{figure}[t]
\includegraphics[width=\columnwidth]{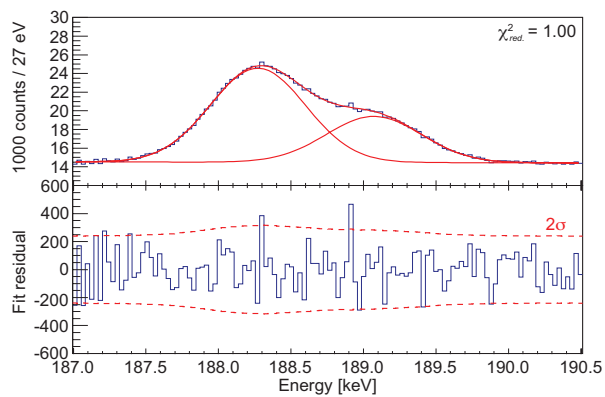}
\caption{(Colour online) \textbf{a)} Part of the $\gamma$-ray singles spectrum detected with the BEGe detector. Indicated is a fit with multiple gaussians with linear background. \textbf{b)} fit residuals with indicated 2$\sigma$ confidence interval (red dashed lines).
\label{fig.Residuals}}
\end{figure}

\begin{figure}[!t]
\includegraphics[width=\columnwidth]{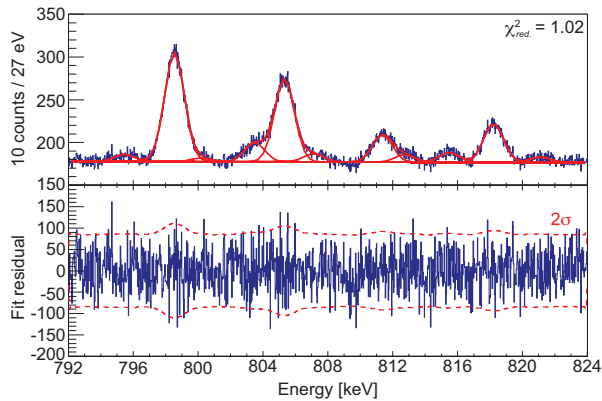}
\caption{(Colour online) \textbf{a)} Part of the $\gamma$-ray singles spectrum detected with the BEGe detector. Indicated is a fit with multiple gaussians with linear background. \textbf{b)} fit residuals with indicated 2$\sigma$ confidence interval (red dashed lines).
\label{fig.ResidualsHigh1}}
\end{figure}

The signals from the HPGe detector preamplifiers were analysed using a commercial Pixie-16, 14-bit, 250\,MHz digitiser \cite{Tan08}, designed by the XIA, Inc. Prior to digitising, the signals were amplified and their dc offsets were adjusted using fast operational amplifiers.  A gain of 8 was used for the BEGe detector, which optimised use of the dynamic range such that signals arising from $\gamma$-rays up to 980\,keV could be digitised. In this mode, each ADC channel was equivalent to approximately 27\,eV.  When combined with the almost ideal gaussian shape~\cite{Mot16} of peaks in the BEGe energy spectrum, this allowed doublets at the level of 0.3\,keV separation to be distinguished by a deconvolution of $\gamma$-ray singles spectra. The data acquisition system was operated in a triggerless mode, i.e., all channels were read out and timestamped individually. For each particular sample of $^{183}$Hg, the internal clock of the digitiser was reset to zero prior to the start of the measurement. The stream of timestamped data was recorded on disk and coincidence information was reconstructed offline.

\section{Level scheme construction}


The level scheme construction procedure used in the present analysis is based on three steps: (i) the determination of $\gamma$ ray energies as precisely as possible using the $\gamma$-ray singles spectrum detected with the BEGe detector, (ii) separation of $\gamma$ rays due to the decay of $^{183}$Hg from its daughter decays and/or room background and (iii) analysis of $\gamma$-$\gamma$ coincidence relationships.  Throughout data acquisition, no instablities in the $\gamma$-ray energy spectra such as peak drift or peak shape degradiation were observed, such that a single set of energy calibration data could be used.  The calibration data were acquired immediately after the end of the $^{183}$Hg samples collection using a $^{152}$Eu source. A quadratic polynomial function was used for the energy calibration of the BEGe detector. In this part of the analysis, the coaxial detectors were not employed.

Due to the excellent charge collection in the BEGe detector and good statistics from the experiment, the peaks in the acquired energy spectra were almost identifical in gaussian shape with a very smooth background ~\cite{Mot16}. Therefore, a gaussian fit with linear background was used to determine the energies of transitions in the $\gamma$-ray singles spectrum. Fits were performed in several short  intervals, which covered the full energy range of the $\gamma$-ray singles spectrum. For each particular fit, its goodness was investigated by using the reduced chi squared ($\chi^2_{red.}$) test and in addition, fit residuals were plotted. An illustrative example of the fit of two $\gamma$ lines is given in Fig.~\ref{fig.Residuals}a. For the fit given in Fig.~\ref{fig.Residuals}a the $\chi^2_{red.}$\,=\,1.00 was obtained. The fit residuals are depicted in Fig.~\ref{fig.Residuals}b. The red dashed lines indicate the 2$\sigma$ confidence interval. The $\chi^2_{red.}$ of 1.00 and fit residuals well within the 2$\sigma$ confidence suggest that fitting function, i.e., two separate gaussians with linear background very well describes the data.  

The ability to detect $\gamma$ rays above 0.5\,MeV with very good resolution and almost ideal gaussian shape is a key factor in our analysis, since it allows the precise determination of high-energy $\gamma$ rays and their exact placement in the level scheme.  Due to the amplification of the BEGe detector preamplifier signals, the highest observed $\gamma$-ray energy was 980\,keV, as discussed previously.  We therefore define our high-energy range as 500\,keV to 980\,keV in this experiment.  An illustrative example of the high-energy $\gamma$-ray singles spectrum is given in Fig.~\ref{fig.ResidualsHigh1}a together with the fit used, which was 11 gaussian distributions with linear background. The $\chi^2_{red.}$ of 1.02 indicates a very good fit, which is corroborated with fit residuals well within 2$\sigma$ confidence interval, as shown in Fig.~\ref{fig.ResidualsHigh1}b.


A typical complication related to odd-mass studies in a region where intruder configurations are present at low excitation energies is a great level density (typically 15-30 excited states below 500\,keV) and thus complexity of measured spectra. This complexity is enhanced by the presence of $\gamma$ rays from other decay branches of the studied isotope, the decay of daughter isotopes, radioactive-ion beam contaminants, interaction of neutrons that are present in the experimental hall and the room background. The  BEGe detector allows the elucidation of a complex decay scheme in the presence of these complicating factors. The data were sorted into a $\gamma$-ray energy vs. timestamp matrix. Subsequently, $\gamma$-ray spectra using 0\,-\,5\,s and 0\,-\,30\,s time windows were projected. The background, which was slightly different in these projections, was subtracted using the TSpectrum class~\cite{Mor97} of the ROOT package. Both spectra were normalised using known $\gamma$ rays due to $^{183}$Hg and $^{183}$Au decays, subsequently subtracted from each other in a way that a spectrum containing only $\gamma$ rays due to the decay of $^{183}$Hg was produced. A separate spectrum dominated by $\gamma$ rays due to the daughter activities, room background, etc., was also produced. This allows for unambiguous isolation of $\gamma$ rays due to the decay of the investigated isotope. 

\begin{figure}[t]
\includegraphics[width=\columnwidth]{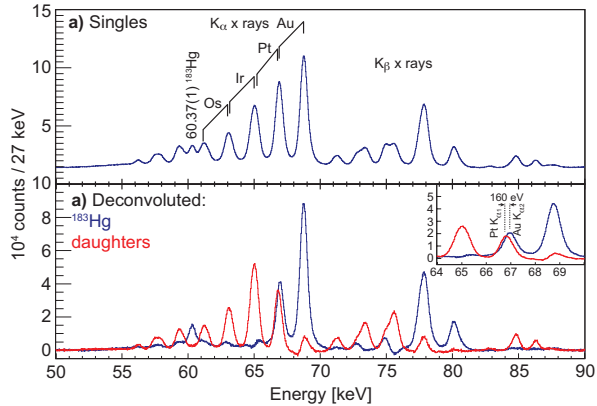}
\caption{(Colour online) \textbf{a)} Spectrum of $\gamma$-ray singles of mass-separated $^{183}$Hg samples. K$_\alpha$ characteristic X rays are evident. Note that, Au K$_{\alpha2}$ (66.991\,keV) and Pt K$_{\alpha1}$ (66.831\,keV) lines are not resolved and same applies for Pt, Ir, Os, etc. \textbf{b)} Deconvoluted spectra of $\gamma$-ray singles of $^{183}$Hg isotope (blue spectrum) and its daughter decays (red spectrum), see the text for details. The insert gives the expansion of both spectra, the Au K$_{\alpha2}$ and Pt K$_{\alpha1}$ lines are clearly resolved.
\label{fig.deconvXrays}}
\end{figure}

The power of this technique is demonstrated in the example shown in Fig.~\ref{fig.deconvXrays}a, which gives the total $\gamma$-ray singles spectrum detected with the BEGe within the 50\,-\,90\,keV energy range. The monotonic sequence of Au, Pt, Ir, and Os K$_\alpha$ characteristic x-rays is distorted only with the presence of the 60.37\,keV peak that is known to occur in the $^{183}$Hg decay~\cite{Mac84}. Fig.~\ref{fig.deconvXrays}b) shows the separated spectra of events due to the $^{183}$Hg decay (blue) and due to daughter decays (red). In the $^{183}$Hg decay spectrum (blue) only Au K$_{\alpha1}$ and K$_{\alpha2}$ are present, while characteristic x-rays of lighter elements, i.e., due to daughter decays, are subtracted. The known 60.37\,keV transition is properly identified to be due to the $^{183}$Hg decay. In the $\gamma$ ray singles spectrum shown in Fig.~\ref{fig.deconvXrays}a, the Au K$_{\alpha2}$ (66.991\,keV) and Pt K$_{\alpha1}$ (66.831\,keV) lines are not resolved. After the deconvolution, both components are clearly resolved, as shown in the Fig.~\ref{fig.deconvXrays}b insert. Similar analysis was performed throughout the full energy range of the BEGe detector $\gamma$ ray spectrum.

\begin{figure*}[t]
\includegraphics[width=0.95\columnwidth]{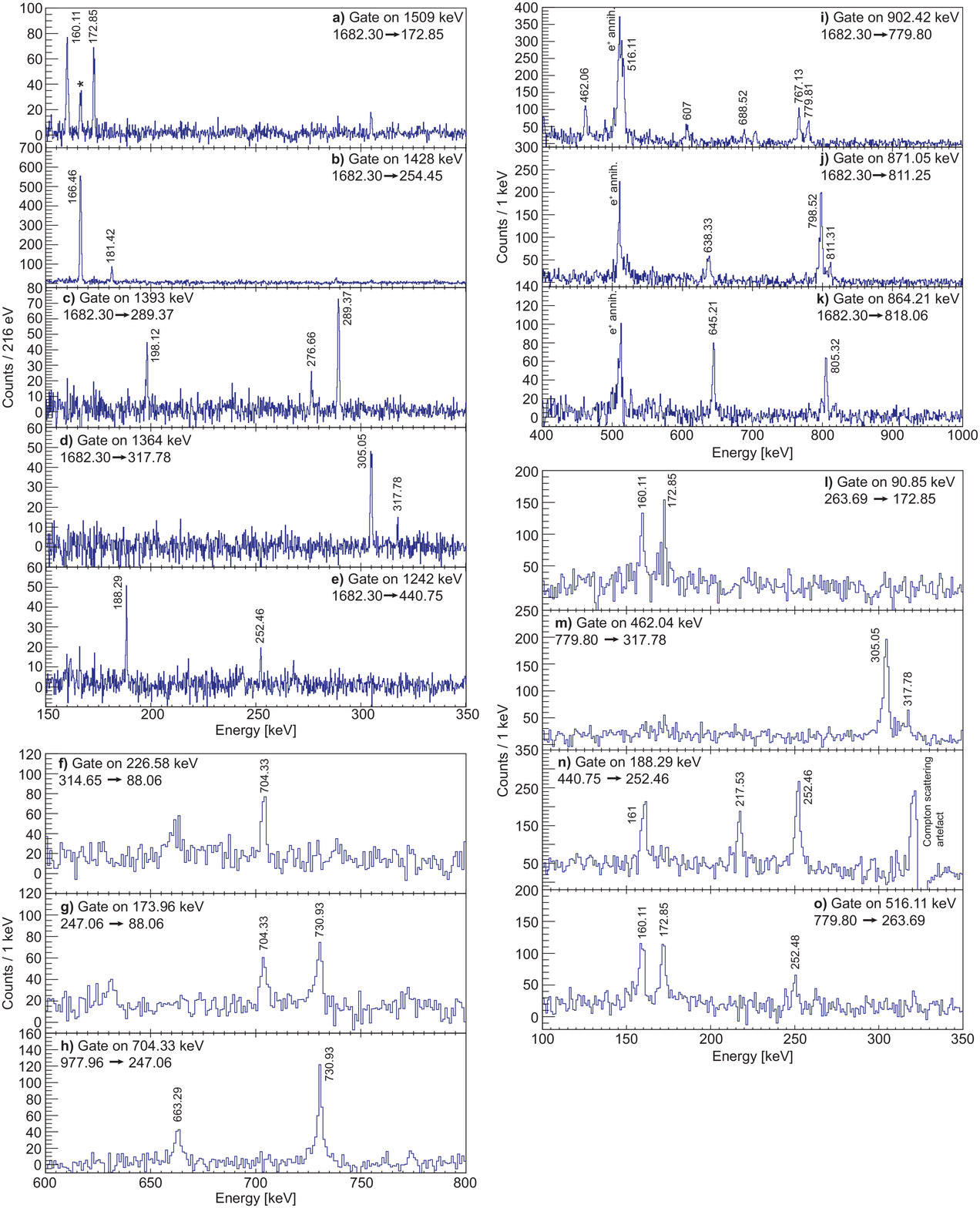}
\caption{(Colour online) Spectra of $\gamma$ rays detected in a prompt coincidence with \textbf{a)} 1509\,keV, \textbf{b)} 1428\,keV, \textbf{c)} 1393\,keV, \textbf{d)} 1364\,keV, \textbf{e)} 1242\,keV, \textbf{f)} 226.58\,keV, \textbf{g)} 173.96\,keV,  \textbf{h)} 704.33\,keV, \textbf{i)} 902\,keV, \textbf{j)} 871\,keV, \textbf{k)} 864.21\,keV, \textbf{l)} 90.85\,keV, \textbf{m)} 462.04\,keV, \textbf{n)} 188.29\,keV, and \textbf{o)} 516.11\,keV $\gamma$ rays in $^{183}$Au. The line marked by a star in the 1509\,keV gate, see panel \textbf{a)}, is the 166.46\,keV $\gamma$ ray and is due to a 1505\,keV $\gamma$ ray in $^{183}$Au (not discussed here). The peak denoted as 'Compton scattering artefact' in panel \textbf{o)} is due to scattering of 511\,keV positron annihilation quanta between the BEGe and coaxial detectors. Note that indicated energies were determined using the $\gamma$-ray singles spectrum measured with the BEGe detector, see the text for details.
\label{fig.GammaGamma}}
\end{figure*}

\begin{figure*}[t]
\includegraphics[width=\columnwidth]{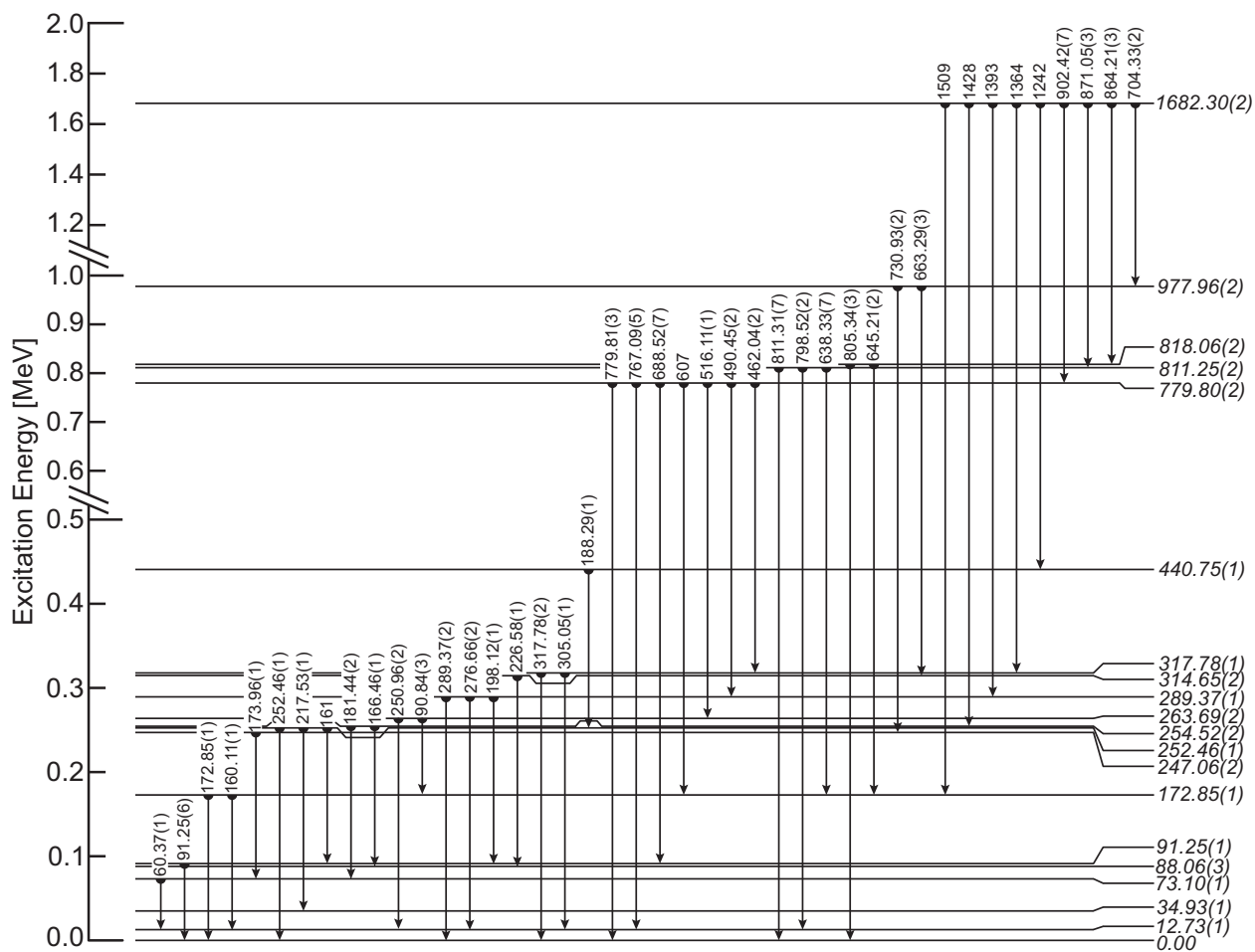}
\caption{Level scheme of $^{183}$Au deduced in present work. Transition energies were determined using the $\gamma$-ray singles spectrum detected with the BEGe detector, see the text for details. Note that $\gamma$ rays above 1\,MeV were detected with coaxial detectors only and their energies could not be determined precisely due to high density of lines, therefore are given only only as integer. Also note that the 161 and 607\,keV transitions were dominated with strong lines arising from daughter activities and therefore could not be determined precisely. However, their placement in the level scheme is evident from $\gamma$-$\gamma$ coincidence analysis.
\label{fig.Scheme}}
\end{figure*}

To construct the level scheme, $\gamma$-$\gamma$ coincidences of three types were analysed separately: (i) gating on the BEGe and projecting spectrum of coaxial detectors, (ii) gating on coaxial detectors and projecting the spectrum of the BEGe detector and (iii) coincidences between both coaxial detectors. To enhance the statistical quality of the coincidence spectra, all available data were used, corresponding to the 32 hours of measurement. Relevant coincidence spectra are given in Fig.~\ref{fig.GammaGamma} and the constructed level scheme in Fig.~\ref{fig.Scheme}.

\begin{figure}[t]
\includegraphics[width=\columnwidth]{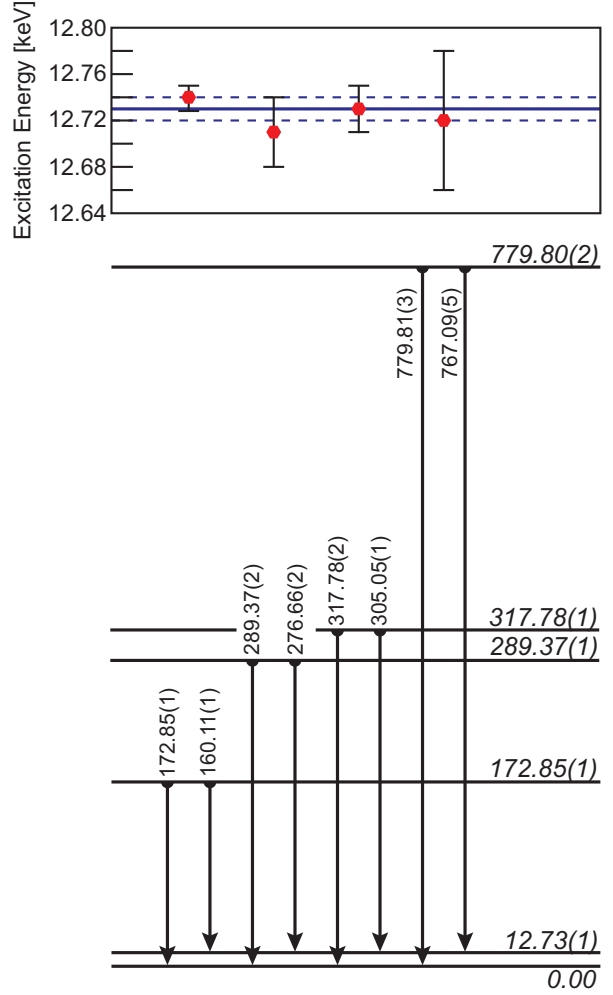}
\caption{(Colour online) Rydberg-Ritz combination principle used to determine the excitation energy of the first excited state. Four differences of $\gamma$-ray feeding the ground state and first excited state were used to determine the precise energy. The difference are given in a graph above the partial level scheme. 
\label{fig.FirstExcitedState}}
\end{figure}

Transitions were localised in the level scheme using the coincidence relationships and Rydberg-Ritz combination principle, facilitated through the precise determination of energies from the BEGe spectrum, see a discussion above). In the case that several combinations, i.e., various sums or differences were available, the excitation energies of  levels were determined using the weighted average. A typical example is the first excited state, in which the excitation energy was determined using energy differences between transitions feeding the ground state and first excited state, arising from deexcitation of the 178.25, 289.37, 317.78, and 779.80\,keV states, shown in Fig.~\ref{fig.FirstExcitedState}. Energy differences of transitions feeding the ground state and first excited state, respectively of 12.74(1), 12.71(3), 12.73(2), and 12.72(5)\,keV were also obtained, as shown in Fig.~\ref{fig.FirstExcitedState}. Within experimental uncertainties all values are consistent. Weighted averaging of these values gives the energy of 12.73(1)\,keV for the first excited state of the $^{183}$Au. This approach has been used for all energy levels indicated in the level scheme given in Fig.~\ref{fig.Scheme}.

\begin{figure}[t]
\includegraphics[width=\columnwidth]{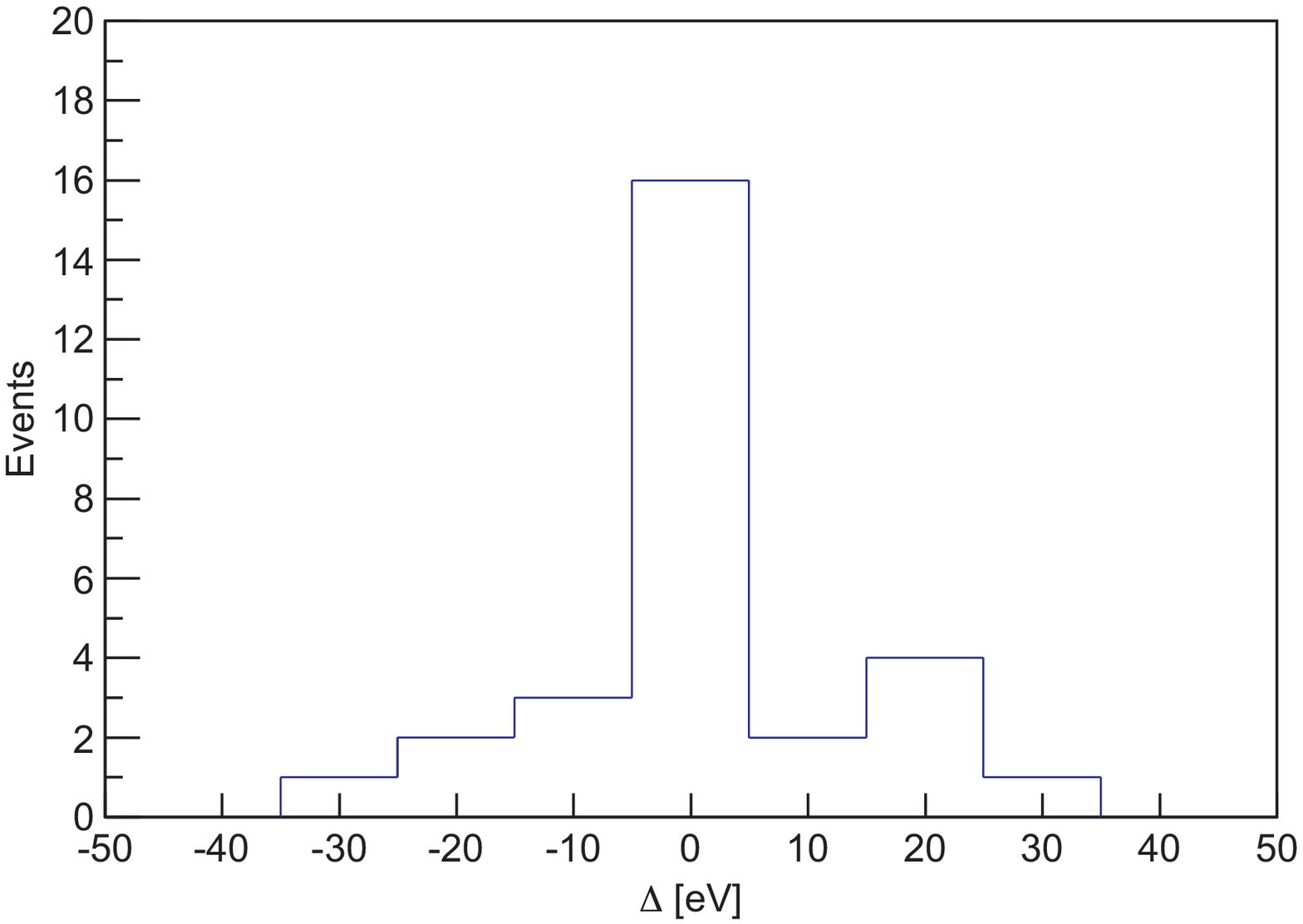}
\caption{(Colour online) Distribution of the $\Delta$, which is a difference between expected transition energy (calculated as a energy difference of initial and final state) and $\gamma$ ray energy detected with the BEGe detector for all relevant events, see Tab.~\ref{tab.GammaRays}.
\label{fig.DistMethod}}
\end{figure}

The excitation energy of the initial and final states together with the $\gamma$ ray energy for identified transitions due to $^{183}$Hg decay are tabulated in Tab.~\ref{tab.GammaRays}. For each transition, the $\gamma$-ray energies obtained using the gaussian fit of the BEGe $\gamma$-ray singles spectrum were compared with expected value, i.e., the difference of excitation energies of initial and final states. The experimental uncertainties were calculated propagating the peak centroid uncertainties and also uncertainties of the coefficients of the calibration polynomial. The difference between the expected and measured values is denoted as $\Delta$ in Tab.~\ref{tab.GammaRays}. A histogram of $\Delta$ values is given in Fig~\ref{fig.DistMethod}. The distribution is centred around 0\,eV. Since only two decimal digits are given for the $\gamma$ ray energies, this means that most of events agree up to 10\,eV. The worst cases are $\pm$\,30\,eV, see Tab.~\ref{tab.GammaRays} and Fig.~\ref{fig.DistMethod}.

\begin{table}
\caption{Excitation energy of initial ($E_i$) and final ($E_f$) states in the $^{183}$Au isotope. Corresponding $\gamma$-ray energy, determined from the $\gamma$-ray singles spectrum detected with the BEGe detector are also given ($E_\gamma$). The $\Delta$ is a difference between expected transition energy, i.e., $E_i$\,-\,$E_f$ and $\gamma$-ray energy ($E_\gamma$). Note that transitions denoted with an asterisk are very weak and on the the present level of the statistics, more precise energy could not be obtained. However, their placement in the level scheme is based on the $\gamma$-$\gamma$ coincidence analysis.
\label{tab.GammaRays}}
\begin{tabular}{cccc}
\hline
\textbf{E$_i$} & \textbf{E$_f$} & \textbf{E$_\gamma$} & $\Delta$\,=\,\textbf{E$_i$}\,-\,\textbf{E$_f$}\,-\,\textbf{E$_\gamma$}	\\  
\textbf{[keV]} & \textbf{[keV]} & \textbf{[keV]} & \textbf{[eV]} 			\\
\hline
73.10(1) & 12.73(1) & 60.37(1) & 0 \\
91.25(1) & 0 & 91.25(6) & 0 \\
172.85(1) & 0  & 172.85(1) & 0 \\
172.85(1) & 12.73(1) & 160.11(1) & -10 \\
252.46(1) & 0 & 252.46(1) & 0 \\
252.46(1) & 34.93(1) & 217.53(1) & 0 \\
254.52(2) & 88.06(3) & 166.46(1) & 0 \\
254.52(2) & 73.10(1) & 181.44(2) & -20 \\
263.69(2) & 12.73(1) & 250.96(2) & 0 \\
263.69(2) & 172.85(1) & 90.84(3) & 0 \\
289.37(1) & 0 & 289.37(2) & 0 \\
289.37(1) & 12.73(1) & 276.66(2) & 20 \\
314.65(2) & 88.06(3) & 226.58(1) & -10 \\
317.78(1) & 0 & 317.78(2) & 0 \\
317.78(1) & 12.73(1) & 305.05(1) & 0 \\
440.75(1) & 252.46(1) & 188.29(1) & 0 \\
779.80(2) & 0 & 779.81(3) & 10 \\
779.80(2) & 12.73(1) & 767.09(5) & 20\\
779.80(2)	& 91.25(6) & 688.52(7)* & -30\\
779.80(2) & 263.69(2) & 516.11(1) & 0 \\
779.80(2) & 289.37(1) & 490.45(2) & 20 \\
779.80(2) & 317.78(1) & 462.04(2) & 20 \\
811.25(2) & 0	& 811.31(7)*	&	60 \\
811.25(2) & 12.73(1) & 798.52(2) & 0 \\
811.25(2) & 172.85(1) & 638.33(7)* & -70 \\
818.06(2) & 12.73(1) & 805.34(3) & 10 \\
818.06(2) & 172.85(1) & 645.21(2) & 0 \\
977.96(2) & 247.06(2) & 730.93(2) & 30 \\
977.96(2) & 314.65(2) & 663.29(3) & -20 \\
1682.3(2) & 779.80(2) & 902.42(8)* & -80 \\
1682.3(2) & 811.25(2) & 871.05(3) & 0 \\
1682.3(2) & 818.06(2) & 864.21(3) & -30 \\
1682.3(2) & 977.96(2) & 704.33(2) & -10 \\
\hline
\end{tabular}
\end{table}

The physics impact of the constructed level scheme derived from this experiment in understanding the nuclear structure of odd-Au isotopes will be discussed in a forthcoming paper~\cite{Ven16}.

\section{Conclusion}
A BE2020 BEGe detector operated at ultra-high gain was successfully used to construct the level scheme of $^{183}$Au, which is a nucleus with a large density of excited states at low energy. The advantage of the BEGe detector is not only excellent resolution but also the ability to detect high energy $\gamma$ rays.  Using this detector, $\gamma$-ray energies with a precision below 50\,eV (in most cases even down to 10\,eV) could be determined. To reach such precision, it is critical to operate the detector at ultra-high gains and also to ensure the stability of the electronics. With precisely determined $\gamma$-ray energies, the Rydberg-Ritz combination principle on the 30\,eV precision could be used, which makes the process of complex level scheme construction much more simple. 

Properties of the BEGe detector, particularly the smooth background continuum, allowed the peaks of interest to be distinguished from other processes such as from the decay of daughter activities. This is very important since it simplifies the analysis and reduces the risk of misinterpreting observed transitions.  It also provides additional information on the daughter isotopes, which can be analysed separately. 

Even with the precision of the BEGe detector, the $\gamma$-$\gamma$ coincidence analysis cannot be omitted. Therefore it appears an optimum solution to combine the BEGe detectors with either larger volume BEGe detectors or conventional coaxial detectors or clovers, which offer higher efficiency for $\gamma$ rays above 1\,MeV. Such systems will play a major role in  the future $\beta$ decay studies of isotopes that have many excited states at low energies.  Other promising detectors are novel Small Anode Germanium detectors (SAGe)~\cite{Ade15}, which combine the good energy resolution the BEGe with better efficiency for $\gamma$ rays above 1\,MeV.

\section*{Acknowledgement}
Authors express their gratitude to the ISOLDE collaboration, ISOLDE machine operators, and CERN radioprotection team for excellent support. Very special thanks goes to the ISOLDE physics coordinator Magdalena Kowalska. This work was supported by the Slovak Research and Development Agency under contract No. APVV-0177-11, by the Slovak Grant Agency VEGA under contract No. 2/0121/14, by STFC UK, by STFC Consolidated Grant No. ST/F012071/1, by STFC Consolidated Grant No. ST/L005670/1, by STFC Continuation Grant No. ST/J000159/1, by EU Seventh Framework through ENSAR No. 506065, by IWAP - Belgian Science Policy (BRix network P7/12), by GOA 10/010 from KU Leuven, and by FWO Flanders. T. E. Cocolios was supported by STFC Ernest Rutherford Fellowship No. ST/J004189/1.
\section*{References}

\bibliography{183Au.bib}

\end{document}